\documentclass[amsmath, amssymb, twocolumn]{revtex4-1}

\usepackage{appendix}
\usepackage[utf8]{inputenc}
\usepackage[section]{placeins}
\usepackage[dvipsnames]{xcolor}
\usepackage{bbold}
\usepackage{graphicx}
\usepackage{dsfont}
\usepackage[hidelinks]{hyperref}
\usepackage{amsmath}

\usepackage{booktabs}
\AtBeginDocument{% this is revtex4 fault
  \heavyrulewidth=.08em
  \lightrulewidth=.05em
  \cmidrulewidth=.03em
  \belowrulesep=.65ex
  \belowbottomsep=0pt
  \aboverulesep=.4ex
  \abovetopsep=0pt
  \cmidrulesep=\doublerulesep
  \cmidrulekern=.5em
  \defaultaddspace=.5em
}

\usepackage{braket}
\usepackage{multirow}
\usepackage{tabularx}

\usepackage{glossaries}
\newacronym{hf}{HF}{Hartree-Fock} % chktex 8
\newacronym{uhf}{UHF}{unrestricted \gls{hf}} % chktex 8
\newacronym{mp2}{MP2}{second-order M\o ller-Plesset Perturbation Theory}
\newacronym{ump2}{UMP2}{unrestricted \gls{mp2}}
\newacronym{CIS}{CIS}{Configuration interaction Singles}
\newacronym{RPA}{RPA}{Random Phase Approximation}
\newacronym{ccsd}{CCSD}{Coupled Cluster Singles Doubles}
\newacronym{uccsd}{UCCSD}{Unrestricted \gls{ccsd}}
\newacronym{ccsd(t)}{CCSD(T)}{perturbative triples} % chktex 36
\newacronym{dft}{DFT}{Density Functional Theory}
\newacronym{tddft}{TD-DFT}{Time-Dependent Density Functional Theory}
\newacronym{cc}{CC}{Coupled Cluster}
\newacronym{fci}{FCI}{Full Configuration Interaction}
\newacronym{eom}{EOM}{Equation Of Motion}
\newacronym{eomcc}{EOM-CC}{Equation of motion Coupled Cluster}
\newacronym{ipeomcc}{IP-EOM-CC}{Ionization Potential \gls{eomcc}}
\newacronym{eaeomcc}{EA-EOM-CC}{Electron Attachment \gls{eomcc}}
\newacronym{eeeomcc}{EE-EOM-CC}{Electron Excitation \gls{eomcc}}
\newacronym{eeeomccsd}{EE-EOM-CCSD}{Equation of motion \gls{ccsd}}
\newacronym{cc4s}{\texttt{cc4s}}{Coupled Cluster For Solids}
\newacronym{ctf}{\texttt{CTF}}{Cyclops Tensor Framework}
\newacronym{bse}{BSE}{Bethe-Salpeter equation}
\newacronym{qd}{QD}{quantum dot}
\newacronym{cbs}{CBS}{complete basis set}
\newacronym{qmc}{QMC}{Quantum Monte Carlo}

\begin{document}

\title{%
  Coupled cluster theory for the ground and excited states of two dimensional
  quantum dots}

\author{Faruk Salihbegovi\'c}
\author{Alejandro Gallo}
\author{Andreas Grüneis}
\affiliation{Institute for Theoretical Physics,
 Vienna University of Technology (TU Wien).
 A-1040 Vienna, Austria, EU.}

\date{\today}

\begin{abstract}
  We present a study of the two dimensional circular quantum dot model
  Hamiltonian using a range of quantum chemical ab initio methods.
  Ground and excited state energies are computed on different levels of
  perturbation theories including the coupled cluster method.
  We outline a scheme to compute the required Coulomb integrals in real space
  and utilize a semi-analytic solution to the integral over the Coulomb kernel
  in the vicinity of the singularity.  Furthermore, we show that the remaining
  basis set incompleteness error for two dimensional quantum dots scales with the
  inverse number of virtual orbitals, allowing us to extrapolate to the complete
  basis set limit energy.
  By varying the harmonic potential parameter we tune the correlation strength
  and investigate the predicted ground and excited state energies.
%NOTE:
%  The ground state energies are compared to quantum Monte
%  Carlo findings and The excitation energies are compared on different levels of
%  perturbation theories.
\end{abstract}

\maketitle
%\tableofcontents

\section{Introduction}

% Rough introduction to quantum dots
%
%
A \gls{qd} is a semiconducting nanocrystal typically embedded in a host
semiconductor with a larger band gap such that the excitons of the \gls{qd}
have a de Broglie wavelength comparable to the size of the crystal.
The typical size of such a nanocrystal is 2~nm--100~nm and it is made out of
roughly a million atoms.
In this context,
virtually all electrons are tightly bound to the nuclei of the material such
that the number of free electrons in a \gls{qd} ranges typically from 1 to 100.
As described by the
quantum mechanical theory of solids, the electrons do not get trapped in the
real nuclei of the material but instead simply sense a potential well of the
\gls{qd}, thus forming discrete energy levels. These electrons behave as free
electrons with a renormalized mass.
For example, electrons in the semiconductor GaAs appear to carry a mass of only
7\% of the mass of free electrons.
\glspl{qd} are often referred to as artificial atoms because they exhibit
similar properties as atoms (level spacing, ionization energy, magnetic moments)
albeit on different energy scales.

% Practical applications of quantum dots
%
%
Due to their tunable optical and electronic properties, \glspl{qd} are
widely used in many practical applications
including
  solar cells,
  light-emitting diodes,
  laser technology as well as
  biological and biomedical
  applications~\cite{ref11,ref12,ref13,6.0,30.0, Bester2003, Singh2009}.
The use of \glspl{qd} as
cosmetic hair dyes is the oldest known application, dating back more than
2000 years, when PbS \glspl{qd} were synthesized using naturally occurring
materials like Ca(OH)$_2$, PbO and water~\cite{ref1}.
Over the past few decades, several ways to synthesize and investigate dynamical properties of \glspl{qd}
with extraordinary high precision have been developed~\cite{ref6}.
Consequently, experimental and theoretical
research on these nanoparticles has harnessed much
attention and insight~\cite{ref7, ref8, ref9, ref10, 8.0, 9.0, 13.0, 14.0, 17.0, 18.0, 19.0, 21.0, 22.0, 23.0, 24.0, 33.0, 34.0, 43.0, 44.0, 45.0, 46.0, 47.0, 48.0, 49.0,  Bester2003, Singh2009}.

% Introduce QD model Hamiltonian
%
%
The simplest model used in theoretical studies of \glspl{qd}, which has proven to be
adequate, is the harmonic oscillator~\cite{ref15}.
In this model, the interaction of the electrons with the surrounding
semiconductor material is approximated through the material-specific effective
mass of the electrons and a material-specific relative dielectric constant
that screens the Coulomb interaction. In passing we note that a more realistic nanoscale model
of \glspl{qd} can be obtained by an empirical pseudopotential based approach~\cite{Wang1999,Bester2008}.
In contrast to other many-body systems, in \glspl{qd},
the coupling strength of the two-body operator
relative to the one-body operator can be freely
varied over a wide range of values, thus giving rise to various regimes of
interelectronic correlation.
% Many body methods
%
%
The simple expression of the \gls{qd} model Hamiltonian allows for a straightforward
application of many-electron methods that have historically been developed for atoms and
crystals.
%To correctly describe the
%spectrum of the \gls{qd}, one has to account for many-body
%correlation effects and the specific form of the one-body confinement.
%While neither of them are easily extracted from experiments, having a reliable
%theory for the correlation effects leaves the specific form of the confinement to be a tuning problem of the potential.
\gls{dft} using approximate density functionals is computationally extremely
efficient and able to account for a large part of the electron correlation.
However, its general applicability is often hindered by
uncontrolled approximations used for constructing the approximate
exchange-correlation functional.
Notwithstanding its drawbacks, one-electron theories such as
\gls{dft} in the Kohn-Sham framework of approximate exchange and correlation
(XC) energy functionals~\cite{ref16,ref17,ref18,ref19} and the
\gls{hf}~\cite{ref21,ref22,ref23,ref24,1.0,4.0,5.0,9.0,10.0,11.0,28.0} approximation often achieve a
qualitatively correct agreement with experiment.
In contrast to \gls{dft} calculations, \gls{fci} investigations of \glspl{qd}
yield exact results for a given basis set and have been applied to \glspl{qd}
in a number of
studies~\cite{ref25,ref26,ref27,ref28,ref29,ref30,ref31,ref32,ref33,ref34,3.0,6.0,11.0,12.0,15.0,16.0,27.0}.
\gls{fci} employs excited Slater determinants to span a many-body wave function
space and through exact diagonalization finds a superposition of Slater
determinants with the lowest energy.
However, since the size of this space grows
combinatorially fast with respect to the number of particles and basis
functions, \gls{fci} is prohibitively computationally expensive.
Alternatively, a technique that has been used for \glspl{qd} thoroughly is
\gls{qmc}~\cite{ref36,ref37,ref38,ref39,ref40,ref41,ref42,ref43,7.0,16.0}.
Here the computational cost grows relatively modestly with the number of
electrons and it provides highly accurate ground state energies.
Moreover, there is the
possibility to use the nodal structure of the ground state trial wave function
to impose restrictions on the solutions.
In this way, excited states can be calculated as well even if
calculations on general excited states are not straightforward.
\gls{cc} theory
combines accuracy with feasibility, being numerically less expensive than \gls{fci}
while having size consistency by construction and providing ground state and
excited state energies with an accuracy that is comparable to quantum \gls{qmc}
calculations~\cite{ref48,ref49}.
\Gls{mp2} and \gls{ccsd} have been shown to be useful approaches to
calculate atomic, molecular and solid-state
properties~\cite{bartlett,%
                 Applying.the.CoGruber2018,%
                 Laplace.transfoSchafe2018,%
                 CeriumOxidesWSchafe2021,%
                 Second.order.MoGrunei2010,32.0}.
They have also been used to study
\gls{qd} Hamiltonians in a number of
studies~\cite{ref44,ref45,ref49,ref45,ref47,ref48}.
Via the \gls{eom} formalism,
\gls{cc} theory can also be applied to excited
states~\cite{The.equation.ofStanto1993}, and was already applied to atoms,
molecules and recently even
solids~\cite{bartlett,%
             Equation.of.MotKrylov2008,%
             Excitons.in.SolWang.2020,%
             Gallo_2021}.
Here, we seek to apply \gls{eeeomccsd} theory to study excited states in two
dimensional \glspl{qd}~\cite{2.0,20.0}.
To this end, we employ an implementation of \gls{eeeomccsd} that was recently
used to investigate defects in solids employing ab initio
Hamiltonians~\cite{Gallo_2021}.

% Paper organization
%
%
This paper is organized as follows.  In section~\ref{SPH} and~\ref{MPH} we
introduce the Hamiltonian of the \gls{qd} and its solutions for the single
particle case.
In section~\ref{ME} we present a way to calculate the Coulomb matrix
elements needed for the many-electron Hamiltonian by employing analytical
solutions for the integral over the Coulomb kernel near the singularity in a 4D
hypercube.
Further in section~\ref{HFAPHFT}
we provide an overview of the \gls{ccsd} and \gls{eeeomccsd}
approaches used to calculate the ground state and excited state energies.
In section~\ref{GSE} and~\ref{EE} we present the results of the \gls{ccsd} and
\gls{eeeomccsd} calculations for the ground state and excited states energies of
\glspl{qd} in different confinement regimes for 2, 6 and 12 electrons and
compare them to other findings from the literature.

\section{Theory and Methods}

\subsection{One-Body Hamiltonian}\label{SPH}

Following the description in Ref.~\cite{ref15}, a \gls{qd} can be modeled as
fermionic particles confined to two dimensions in a parabolic potential.
The corresponding one-body Hamiltonian in such a potential is given
in atomic units by

\begin{gather}\label{singleParticleHamiltonian}
\hat{H}(x,y)
  = \frac{p^2_x + p^2_y}{2}
  + \frac{1}{2}\omega^2\left(x^2 + y^2\right).
\end{gather}

Here, $\omega$ is a measure of the confinement strength of the electron in the
parabolic potential well.
Consequently, the Schr\"odinger equation can be separated into $x$
and $y$ coordinates, resulting in the differential equation for the 1D harmonic
oscillator
\begin{gather}
  \left(
    -\frac{1}{2}\frac{\partial^2}{\partial x^2}
    + \frac{\omega^2 x^2}{2}
  \right)
  \psi(x)
  =
  E\psi(x),
\end{gather}
admitting the well-known solutions
\begin{align*}
\psi_{n}(x)
  &= \frac{1}{\sqrt{2^n n!}}
     \left(
       \frac{\omega}{\pi}
     \right)^{\frac{1}{4}}
     e^{-\frac{\omega x^2}{2}}
     H_n\left(
       \sqrt{\omega}x
     \right) \\
E_n
  &= \omega \left(n+\frac{1}{2}\right) \\
H_n(x)
  &= (-1)^n e^{x^2} \frac{\mathrm d^n}{\mathrm dx^n}\left(e^{-x^2}\right)
\end{align*}
where \( n \in \mathbb{N} \).
The corresponding solutions to the non-interacting 2D problem are
\begin{align}
  \psi_{nm}(x,y)
    &= \psi_n(x)\psi_m(y)\label{2D orbitals} \\
  E_{nm} &=
    E_n + E_m = \omega (n + m + 1)\label{2D energy}.
\end{align}
We note that the ground state is given by the solution with $m = 0$ and $n = 0$
and is non-degenerate.
The first excited state is given by the solutions with $(n=1, m=0)$
and $(n=0, m=1)$ and has a degeneracy of 2.
The second excited state is given by ($n=2$, $m=0$), ($n=0$, $m=2$) and ($n=1$,
$m=1$) and has a degeneracy of 3, and so forth.

\subsection{Two-Body Hamiltonian}\label{MPH}

The electronic structure of the 2D \gls{qd} is strongly affected by electronic
correlation effects caused by inter electronic interactions.  To describe the
true many-body nature of the  2D \gls{qd} with $N$ electrons, we have to include
the two-body Coulomb interaction and consider the following two-body
Hamiltonian
\begin{gather}
  \hat{H}
  = \sum_{i=1}^{N} \hat{H} (x_i,y_i)
  + \frac{1}{2}\sum_{i \neq j}^N
    \frac{1}{\sqrt{(x_{i}-x_{j})^2+(y_{i}-y_{j})^2}},
\end{gather}
where $\hat{H} (x_i,y_i)$ is the one-body operator defined by equation
\ref{singleParticleHamiltonian}.
Herein we employ the bare Coulomb interaction. We note, however, that many model
Hamiltonians for \glspl{qd} account for screening effects by including various
approximations to the permittivity in the inter electronic interaction.
Given the fermionic character of the particles, the form and relative strength
of the one-particle and two-particle operators of the above Hamiltonian, it is
reasonable to assume that conventional quantum chemical many-electron wave
function based methods yield reliable solutions for its ground and excited
states.
In this hierarchy of quantum chemical wave function based methods, the
\gls{hf} theory, employing a self consistent field approximation, is a
well-established starting point.

\subsection{\gls{hf} and post-\gls{hf} theory}\label{HFAPHFT}

The \gls{hf} method is one of the simplest wave function based
ab initio
approaches used in electronic structure theory calculations.
It serves not only as a useful approximation in its own right but as the
starting point of other more accurate models such as \gls{cc}.
In \gls{uhf} theory the many-body wave function is approximated by a single
Slater determinant and the energy is optimized with respect to variations of the
spin orbitals used to construct the Slater determinant.
The Slater determinant formed from these spin orbitals is the \gls{uhf}
ground state wave function $\ket{0}$ and can be interpreted as a new
vacuum from where particle-hole pairs are created and annihilated
in the context of quantum field theory.

Building on one-body theories such as \gls{uhf},
coupled cluster theory employs an exponential ansatz acting on a single Slater
determinant. Using $ \ket{0} $ the ansatz reads
\begin{gather}
  \ket{\Psi_{\mathrm{CC}}}
    = e^{\hat{T}}\ket{0}%
    ,\\
  \hat{T}
    = \sum_{i,a}
        t_i^a \hat{a}_{a}^{\dagger} \hat{a}_i
      + \frac{1}{4}\sum_{i,j,a,b} t_{ij}^{ab}
          \hat{a}_{a}^{\dagger} \hat{a}_{b}^{\dagger} \hat{a}_j\hat{a}_i
      + \cdots
\end{gather}
where indices $a, b, \ldots$ and $i, j, \ldots$
denote virtual or particle and occupied or hole
orbitals, respectively.  $\hat{a}^{\dagger}$ and $\hat{a}$ are the second
quantization creation and annihilation operators, creating excited Slater
determinants when acting on the reference determinant.
The cluster operator $ \hat{T} $ includes in principle all excitations
up to the number of electrons in the system.
Using this exponential form for the wave function ansatz in the stationary
Schr\"{o}dinger equation gives
\begin{align*}
  \hat{H} \ket{\Psi_{\mathrm{CC}}}
    &= E_{\mathrm{CC}} \ket{\Psi_{\mathrm{CC}}},
\end{align*}
which is equivalent to
\begin{align}
  \bar{H} \ket{0}
    &= E_{\mathrm{CC}} \ket{0} \label{CCSD equ},
\end{align}
where we have implicitly defined the similarity transformed Hamiltonian
\( \bar{H} = e^{-\hat{T}} \hat{H} e^{\hat{T}} \).
In coupled cluster singles and doubles theory, the cluster operator $\hat T$ is
truncated such that it includes only singles and doubles excitations.
In order to solve equation~\ref{CCSD equ} for \gls{ccsd} theory,
one has to find the coefficients $t_i^a$ and $t_{ij}^{ab}$.
Working equations are
commonly obtained by projecting the \gls{hf}, singles and doubles manifold
basis $
\left \{
  \ket{0},
  \hat{a}_{a}^{\dagger} \hat{a}_i\ket{0},
  \hat{a}_{a}^{\dagger} \hat{a}_{b}^{\dagger} \hat{a}_j\hat{a}_i\ket{0}
\right \}
$
of the Slater determinant space onto equation \ref{CCSD equ}:
\begin{align}
  E_{\mathrm{CC}}
    &= \bra{0}\bar{H}\ket{0},
      \label{ccsd pr1} \\
  0
    &= \bra{0}
        \hat{a}^{\dagger}_i\hat{a}_a\bar{H}
      \ket{0},
      \label{ccsd pr2}\\
  0
    &= \bra{0}
        \hat{a}^{\dagger}_i\hat{a}^{\dagger}_j\hat{a}_b\hat{a}_a\bar{H}
      \ket{0}.
      \label{ccsd pr3}
\end{align}
Equation~\ref{ccsd pr1} gives an expression for the \gls{cc} energy
and is valid also for non-truncated cluster operators $ \hat{T} $.
Equations~\ref{ccsd pr2}-\ref{ccsd pr3}
form a set of coupled non-linear equations and can be
solved for $t_i^a$ and $t_{ij}^{ab}$ using iterative
methods~\cite{bartlett,book:293288}.

A successful method of obtaining excited states from \gls{cc}
theory is to diagonalize the Hamiltonian $\bar{H}$ in a suitable subspace of the
Hilbert space.
This is the main approach followed in the \gls{eomcc} method.
In this work, we use the charge-neutral variant of this methodology,
the \gls{eeeomcc}
theory~\cite{The.equation.ofStanto1993,Equation.of.MotKrylov2008}.
The theory is based on a linear ansatz for the excitation operators
$ \hat{R} $ as in \gls{fci}, thus the main working equations read
\begin{gather}
  \hat{H}\hat{R}\ket{\Psi_{\mathrm{CC}}}
    = E_R
      \hat{R}\ket{\Psi_{\mathrm{CC}}}
      , \label{eom equ} \\
  \hat{R}
    = r_0
    + \sum_{i,a}
        r_i^a
        \hat{a}_{a}^{\dagger}
        \hat{a}_i
    + \frac{1}{4}\sum_{i,j,a,b}
        r_{ij}^{ab}
        \hat{a}_{a}^{\dagger}
        \hat{a}_{b}^{\dagger}
        \hat{a}_j
        \hat{a}_i
    + \ldots
\end{gather}
The scalars $ \{r_0, r^a_i, r^{ab}_{ij}, \ldots \} $
define the excitation operator $ \hat{R} $ and are to be determined,
whereas $E_{R}$ is its excitation energy.
Equation~\ref{eom equ} is equivalent to a commutator equation only involving
$\bar{H}$ and the excitation energy difference $\Delta E_R$ between $E_R$ and
$E_{\mathrm{CC}}$.
\begin{gather}
  [\bar{H},\hat{R}]\ket{0}
    =
      (E_R - E_{\mathrm{CC}})
      \hat{R}\ket{0}
\end{gather}
Note that in the commutator on the left hand-side only connected diagrams need
to be considered in the CI expansion.
In this work we use the spin-flip version
of \gls{eomcc}~\cite{krylovSF2006,Equation.of.MotKrylov2008},
whereby no spin-conserving restrictions are imposed to the
$ r^{a}_i, r^{ab}_{ij}, \ldots $ amplitudes.
Moreover, all excited state calculations are performed employing the
\gls{eeeomccsd} approach, where only up to two-body excitation operators
are considered.

\subsection{Matrix Elements}\label{ME}

In order to apply \gls{ccsd} theory to find approximate solutions for the ground
state of the 2D \gls{qd} model Hamiltonian represented in a given orbital basis,
one has to compute the Coulomb Integrals
\begin{widetext}
\begin{equation}\label{Coulomb Integral}
  \int\limits_{-\infty}^{\infty}
  \int\limits_{-\infty}^{\infty}
  \int\limits_{-\infty}^{\infty}
  \int\limits_{-\infty}^{\infty}
    \mathrm{d}x_1
    \mathrm{d}x_2
    \mathrm{d}y_1
    \mathrm{d}y_2
      \frac{%
        \psi_{nm}^*(x_1,y_1)
        \psi_{op}^*(x_2,y_2)
        \psi_{qr}  (x_1,y_1)
        \psi_{st}  (x_2,y_2)
      }{%
        \sqrt{ (x_1-x_2)^2
             + (y_1-y_2)^2
             }
      },
\end{equation}
\end{widetext}
where $ \psi_{nm} $ are the two dimensional orbitals introduced in
equation~\ref{2D orbitals}.
For Gaussian based basis sets and their derivatives,
methods for analytical computation of such integrals exist,
which are commonly based on recursive relations
and can be implemented
on a computer using code generation
facilities~\cite{boys,%
                 Efficient.recurObara.1986,%
                 A.simple.algebrAhlric2006}.

In this work we present a numerical approach
for Coulomb integrals evaluation that is computationally less
efficient but can be applied to arbitrary orbitals. This can potentially be
useful for model Hamiltonians represented in a set of basis functions that are
difficult to expand using Gaussian functions or their derivatives but can be
well represented on a sufficiently dense spatial grid.

The main idea of our approach is to assume that the singular Coulomb kernel
exhibits a more rapid spatial variation than the orbitals and that the employed
real space grid
is dense enough to approximate the orbitals by a constant inside any volume/area
sampled by the grid.  Based on this premise, we discretize the wave function and
integrate over discretized hypercubic volume elements with an edge length of $\Delta x$.
The way in which this was done is described in detail in appendix
\ref{Analytic Solution}.
The resulting numerical expression for the Coulomb integral~\eqref{Coulomb
Integral} can be rewritten as a sum
\begin{widetext}
\begin{gather}
  \sum_{ijkl}
    \psi_{nm}^*(x_i,y_j)
    \psi_{op}^*(x_k,y_l)
    \psi_{qr}(x_i,y_j)
    \psi_{st}(x_k,y_l)
    a_{i-k, j-l}
    \Delta x^3.
\end{gather}
\end{widetext}
where $a_{ij}$ is a system independent matrix that does not depend on $\Delta x$.
$ \left \{ i, j, k, l \right \} $ are here discretization indices
and are not to be confused with hole indices.
Note that the factor $\Delta x^3$ implies that $a_{ij}$ is dimensionless.
Additionally, we reiterate that although this approach is computationally
significantly less efficient than the recursive scheme,
the computational bottle neck in the present study remains in the
\gls{eeeomccsd} calculations.

\section{Results}\label{Results}

We study \glspl{qd} for a range of electron numbers and $\omega
\in \{1.0, 0.5, 0.28\}$.
$\omega$ characterizes the correlation strength in the system
relative to the potential energy.  Large $\omega$ correspond to weakly correlated
systems whereas small $\omega$ correspond to stronger correlated systems~\cite{ref48}.

Throughout this section all quantities are presented in atomic units
(a.u.). In particular, all energy values are therefore given in Hartree.

\subsection{Ground state Energies}\label{GSE}

We first discuss the numerical reliability of our approach.
Let us note that our approach employs a single computational parameter, $\Delta x$,
which defines the grid spacing used for the
real space representation of all orbitals and the numerical integration.
Table \ref{comparisonQMC} shows the computed \gls{ccsd} energies of the two electron system
with $\omega=1.0$ for a range of $\Delta x$.
Our findings show that $300\times 300$ grid points suffice to achieve sub-mHa precision.
For all further Coulomb integral calculations, we have therefore discretized the wave function into
squares of edge length $\Delta x = 0.0342$ a.u.,  in a range where $|\psi(x,y)|^{2} > 10^{-10}$.
However, we note that a careful comparison between results summarized in our work and Refs.~\cite{ref48,2.0}
reveals that the published \gls{ccsd} ground state energies do not always agree to within mHa.
We attribute these discrepancies to different choices of basis sets in the \gls{ccsd}
and Hartree--Fock calculations, which can result in a different convergence behaviour of the
energies to the complete basis set limit. Our basis set extrapolation approach will be discussed in the
following paragraphs.

\begin{table}
\caption{%
  Summary of the convergence of \gls{ccsd} energies for the $N = 2$
  electron system with $\omega = 1.0$ as a function of the number of grid points
  $N_{\mathrm{g}}$ used to represent the wave function.
  The \gls{ccsd} energies have been computed for a finite basis set corresponding
  to 6 oscillator shells or 21 orbitals.
  All energies are in Hartree.
  \label{comparisonQMC}}
\begin{tabularx}{0.5\textwidth}{XXX}
\toprule
  $\Delta x$ (a.u.)
  & $N_{\mathrm{g}}$
  & \gls{ccsd}
  \\
\midrule
0.1025    &100$\times$100   &3.013673\\
0.0513    &200$\times$200   &3.013621\\
0.0342    &300$\times$300   &3.013613\\
0.0256    &400$\times$400   &3.013610\\
0.0205    &500$\times$500   &3.013612\\
0.0171    &600$\times$600   &3.013613\\
%\midrule
%0.5 & 2  \\
%   &6  &11.8112  &11.7888  &11.8055  &11.7837\\
%   &12  &39.2292  &39.159    &39.2194  &39.1516\\
%\midrule
%0.28 & 6  &7.6289   &7.6001    &7.6252    &7.6006\\
%     & 12  &25.7178   &25.6356  &25.7089  &25.6324\\
\bottomrule
\end{tabularx}
\end{table}

\begin{table}
\caption{
  HF energy and correlation energy contributions on the level of \gls{mp2} and \gls{ccsd} theory
  in Hartree for 2 electrons.
  $ N_{\mathrm v} $ denotes the number of virtual orbitals, with
  its value at $ \infty $ being the extrapolated value.
  Our results show that as $\omega$ increases, the HF ground state energies
  increases linearly with $\omega$.
  HF is a good approximation in the limit of large $\omega$ where the inter
  electronic interaction is small compared to the one-body interaction.
  \label{table1}
}
\begin{tabularx}{0.5\textwidth}{XXXXX}
\toprule
    $\omega$ (a.u.)
  & $N_{\mathrm{v}}$
  & \gls{hf}
  & \gls{mp2}
  & \gls{ccsd}\\
\midrule
1.0 & 9  &3.1626    &-0.1182   &-0.1374 \\
    & 14  &3.1618    &-0.1284   &-0.1442 \\
    & 20  &3.1618    &-0.1347   &-0.1482 \\
    & 27  &3.1618    &-0.1395   &-0.1508 \\
    & 35  &3.1618    &-0.1431   &-0.1526 \\
    & 44  &3.1618    &-0.146    &-0.1539 \\
    & 54  &3.1618    &-0.1483   &-0.1548 \\
    & 65  &3.1618    &-0.1501   &-0.1556 \\
    & 77  &3.1618    &-0.1517   &-0.1562 \\
    & $\infty$     &3.1617    &-0.1602  &-0.1596 \\
\midrule
0.5 &9   &1.7998    &-0.107    &-0.1259 \\
    &14  &1.7997    &-0.1149   &-0.1302 \\
    &20  &1.7997    &-0.1202   &-0.1324 \\
    &27  &1.7997    &-0.1242   &-0.1339 \\
    &35  &1.7997    &-0.1272   &-0.1348 \\
    &44  &1.7997    &-0.1296   &-0.1356 \\
    &54  &1.7997    &-0.1315   &-0.1361 \\
    &65  &1.7996    &-0.1331   &-0.1365 \\
    &77  &1.7996    &-0.1345   &-0.1369 \\
    &$\infty$     &1.7997    &-0.1418  &-0.1387 \\
\midrule
0.28 &9   &1.1417    &-0.0962   &-0.1129 \\
     &14  &1.1417    &-0.102    &-0.1151 \\
     &20  &1.1417    &-0.1065   &-0.1162 \\
     &27  &1.1417    &-0.1098   &-0.1169 \\
     &35  &1.1417    &-0.1123   &-0.1174 \\
     &44  &1.1417    &-0.1144   &-0.1178 \\
     &54  &1.1417     &-0.1161   &-0.1181 \\
     &65  &1.1417    &-0.1174   &-0.1183 \\
     &77  &1.1417     &-0.1186   &-0.1185 \\
     &$\infty$     &1.1417    &-0.1249  &-0.1194 \\
\bottomrule
\end{tabularx}
\end{table}

Having assessed the reliability of our numerical approach, we now turn to the
discussion of the ground state results obtained on the level of
\gls{hf}, \gls{mp2} and \gls{ccsd} theories.
We stress that it is necessary to converge all post-\gls{hf}
correlation energies with respect to the employed orbital basis set.
For 3-dimensional ab initio systems and the uniform electron
gas~\cite{Shepherd2012}, it is known from second-order perturbation theory that
the basis set error scales as $1/N_{\mathrm{v}}$, where $N_{\mathrm{v}} $
refers to the number of virtual orbitals. The complete basis set limit
is obtained by extrapolating $N_{\mathrm{v}} \rightarrow \infty$.

\begin{figure}[htb]
\begin{center}
\includegraphics[width=0.5\textwidth]{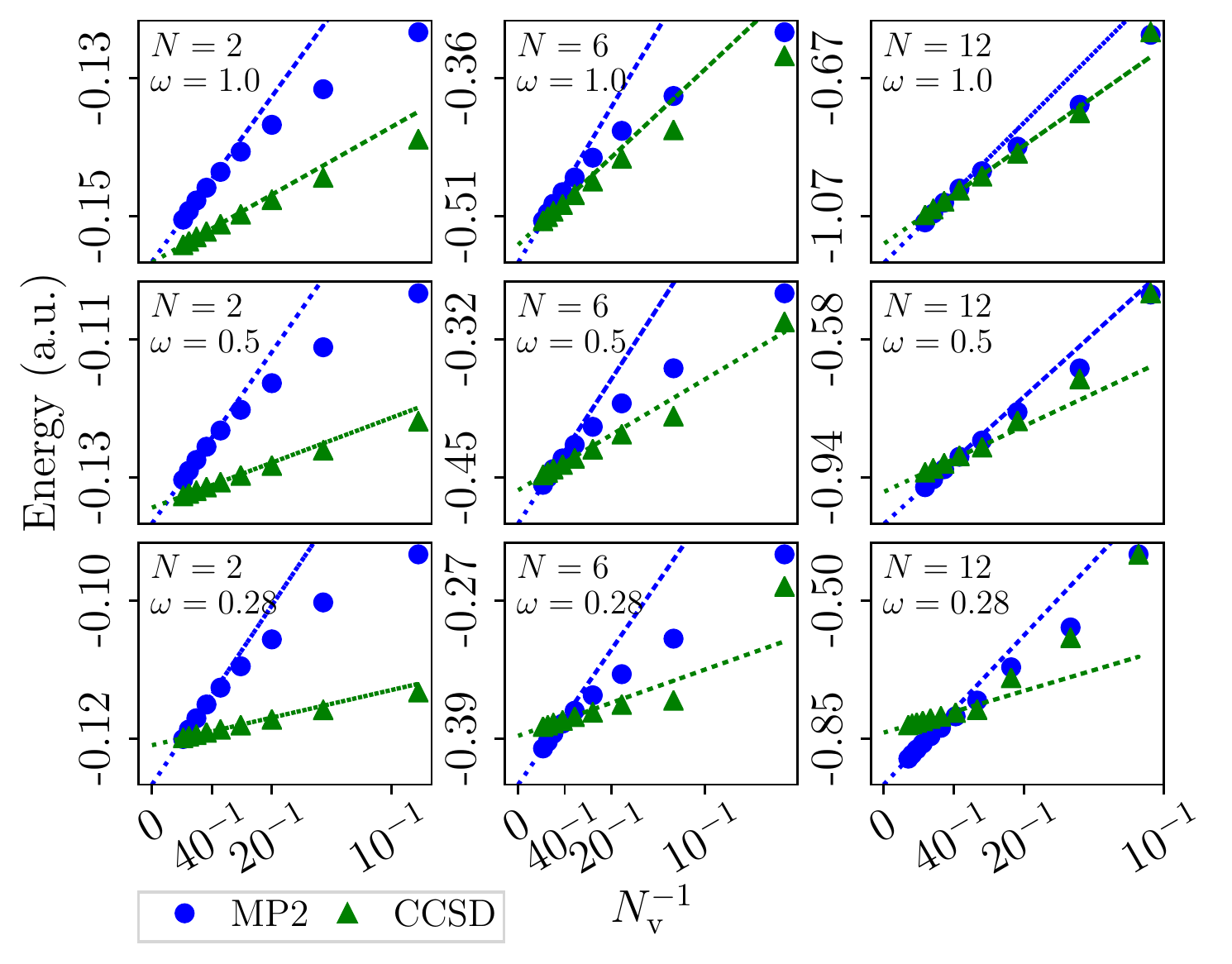}
\caption{%
  \gls{ccsd} and \gls{mp2} correlation energies
  and their \gls{cbs} extrapolations for $N \in \{ 2, 6, 12 \}$ electron
  systems with $\omega \in \{ 1.0, 0.5, 0.28 \} $ as a function of the inverse
  number of virtual orbitals.
  All energies are presented in Hartree and \( \omega \) is given in atomic
  units.
  \label{fig:ccsd and mp2}}
\end{center}
\end{figure}

For the studied system, the two dimensional \gls{qd}, we expect a similar
behavior for the correlation energies.  In order to motivate the validity of
this assumption, we have shown analytically that the asymptotic relation holds
for the second-order perturbation theory correlation energy.
Details regarding the derivation can be found in the appendix~\ref{ABOTCE}.
Numerical results for the correlation energies retrieved as a function of
1/$N_{\mathrm{v}}$ are depicted in Fig.\ref{fig:ccsd and mp2}
and confirm this behaviour for both \gls{mp2} and \gls{ccsd}.
From these numerical findings we conclude that the correlation energies can be
linearly fitted using the following formula
$E(N_{\mathrm{v}}) = E_{\mathrm{CBS}} + \frac{A}{N_{\mathrm{v}}}$
with parameters ($E_{\mathrm{\mathrm{CBS}}}, A)$.
Throughout this work $E_{\mathrm{CBS}}$ refers to extrapolated complete basis
set limit energies that have been obtained by fitting the latter function using
energies obtained with 65 and 77 orbitals.
We estimate that the remaining basis set error is about 0.025\% in the
worst case ($N=12$, $\omega =0.28$~a.u.) and otherwise 0.003~\%.
This estimate corresponds to an approximation and has been obtained by varying
the range of basis set sizes used in the extrapolation

Table~\ref{table1} summarizes the \gls{hf}, \gls{mp2} and \gls{ccsd}
correlation energies together with the \gls{cbs} limit for $\omega \in
\{1.0, 0.5, 0.28\}$ for the 2 electron system.
Compared to the \gls{hf} energy, the \gls{mp2} correlation energy changes only
slightly with $\omega$.
However, on a relative scale the importance of the correlation energy
contribution to the ground state energy increases from 5.3~\% to 11.7~\%
(ratio of \gls{mp2}/\gls{ccsd} correlation energy and the ground state energy
for $N=2$, $\omega=$ 1.0 a.u. and 0.28 a.u.).
Low-order perturbation theories like \gls{mp2} become less reliable in the
regime of strong correlation.
\gls{ccsd}, being a more accurate theory in the sense that it contains all
contributions from MP3 theory and beyond, is expected to yield more
accurate results than \gls{mp2} for small $\omega$.
We can see that the relative \gls{ccsd} and \gls{mp2} contributions to the ground
state energy differ more as $\omega$ decreases.

The linear scaling of the correlation energies with $\omega$ and
the basis set convergence shown in Fig.\ref{fig:ccsd and mp2} is found to 
be qualitatively independent of the number of electrons. 
All calculated \gls{cbs} ground state energies are summarized in 
Table~\ref{cbsgs} for further reference. Our findings demonstrate that small 
electron numbers already serve as a good approximation to the behavior of 
ground state energies with the investigated parameters.

\begin{table}
\caption{%
  Summary of CBS limit CCSD energies for $N \in \{2, 6, 12\}$ electron systems
  with $\omega \in \{ 1.0, 0.5, 0.28 \}$. All energies are in Hartree.
  \label{cbsgs}}
\begin{tabularx}{0.5\textwidth}{XXXXXX}
\toprule
$\omega$ (a.u.) & Electrons	&$E_{\mathrm{CBS}}$	\\
\midrule
1.0 &2	&3.0022		\\
    &6	&20.1839	\\
    &12	&65.7644	\\
\midrule
0.5 & 2	&1.6609		\\
    &6	&11.8118	\\
    &12	&39.2343	\\
\midrule
0.28 & 2  &1.0222 \\
     & 6	&7.6292 \\
     & 12	&25.7190\\
\bottomrule
\end{tabularx}
\end{table}

\subsection{Excitation Energies}\label{EE}

Having established a procedure to converge the ground state energies with the
basis set, we now seek to discuss the excited state properties.
To this end, we employ \gls{eeeomccsd} theory and the same Hamiltonian employed in the
previous section.
We have calculated the first three excitation energies, where the first excited state in \gls{eeeomccsd} theory corresponds to a triplet state while the second and third excited states are singlet states.

\begin{figure}[htb]\begin{center}
\includegraphics[width=0.5\textwidth]{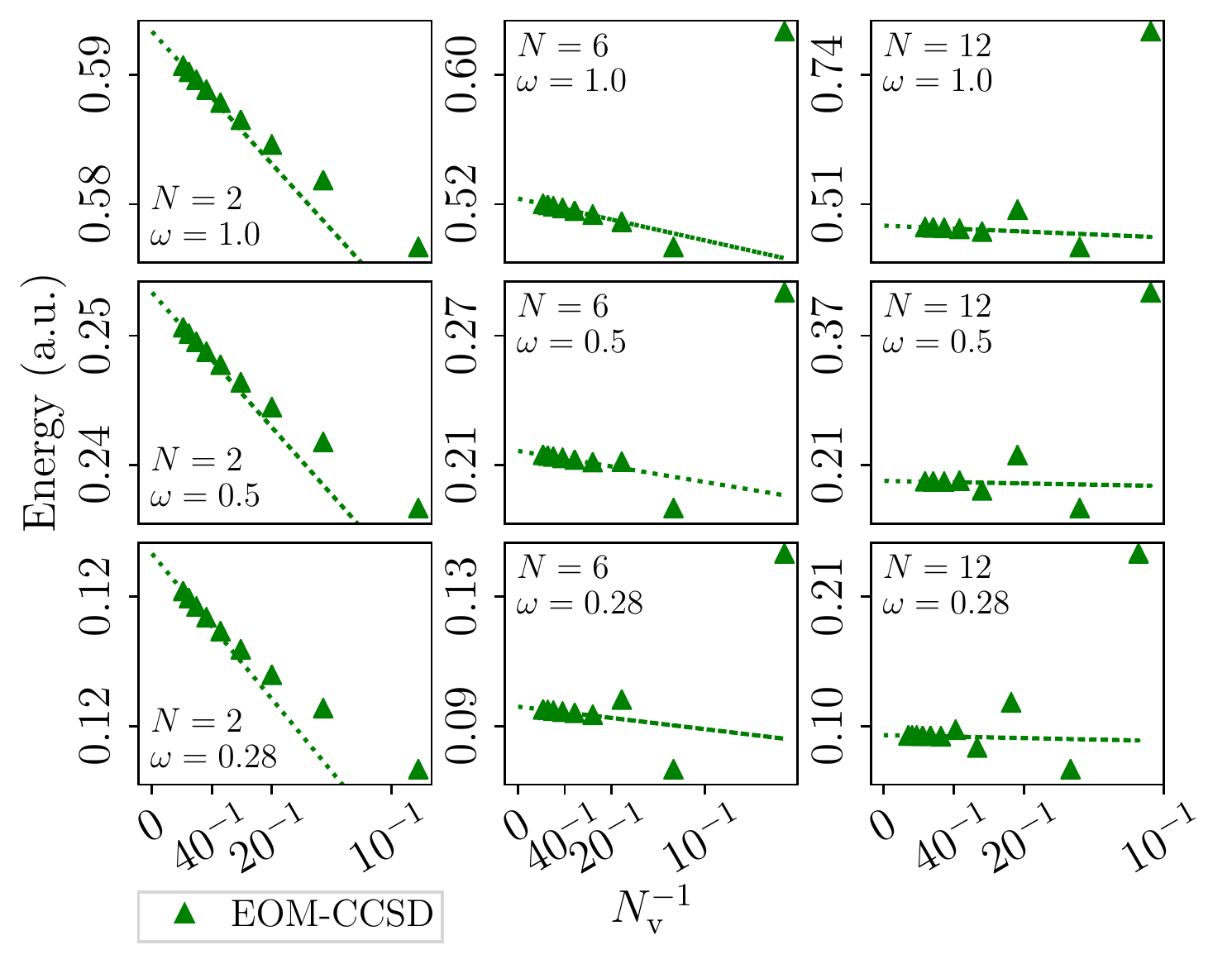}
\caption{%
  First \gls{eeeomccsd} excitation energy for $N \in \{ 2, 6, 12 \}$
  electron systems with $\omega \in \{ 1.0, 0.5, 0.28\}$ retrieved as a
  function of the inverse number of virtual orbitals $N_{\mathrm{v}}^{-1}$.
  All energies are in Hartree.%
}\label{fig:eom_fig}
\end{center}
\end{figure}

\begin{figure}[htb]\begin{center}
\includegraphics[width=0.5\textwidth]{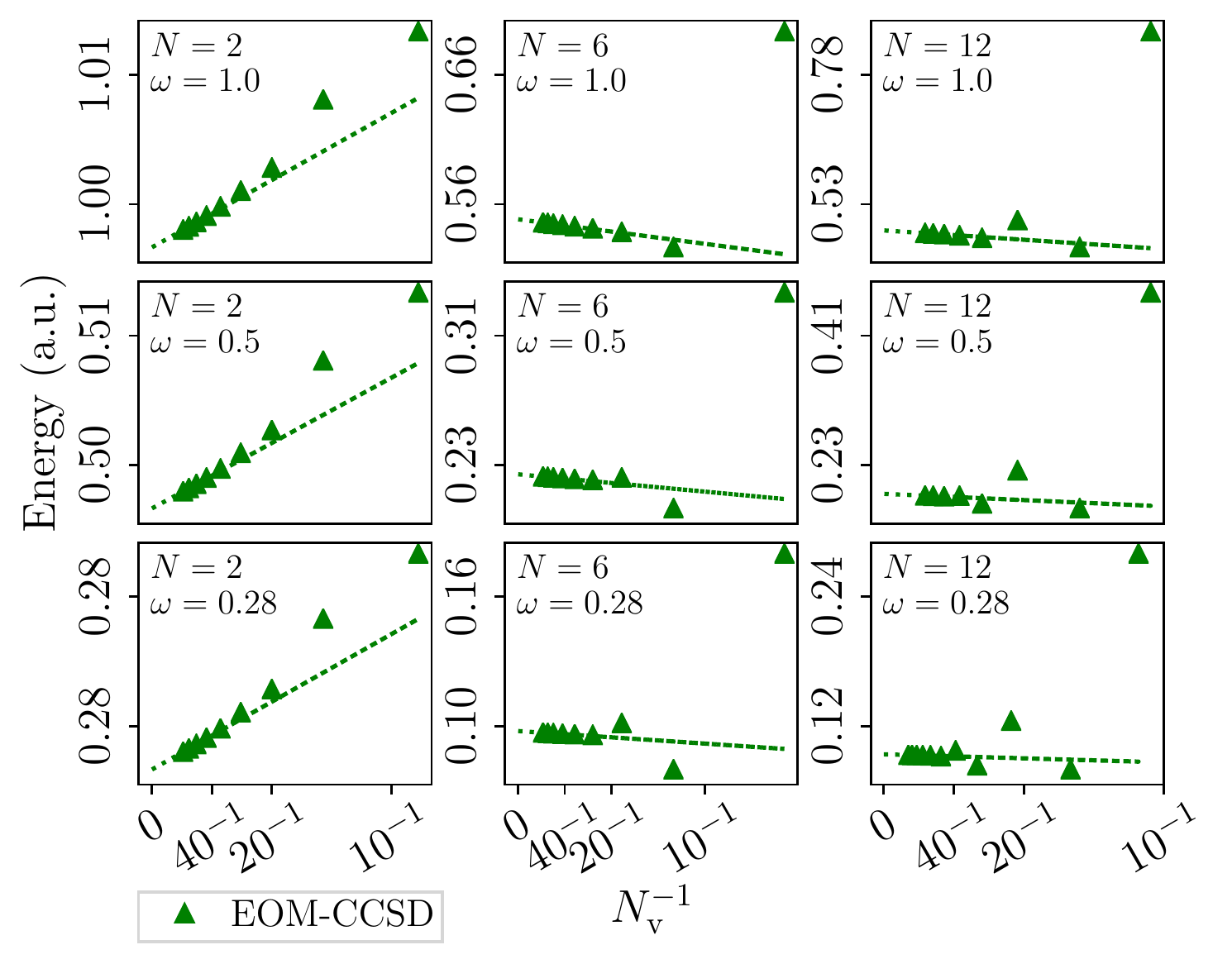}
\caption{%
  Second \gls{eeeomccsd} excitation energy for $N \in \{ 2, 6, 12 \}$
  electron systems with $\omega \in \{ 1.0, 0.5, 0.28\}$ retrieved as a
  function of the inverse number of virtual orbitals $N_{\mathrm{v}}^{-1}$.
  All energies are in Hartree.%
}\label{fig:eom_fig2}
\end{center}
\end{figure}

\begin{figure}[htb]\begin{center}
\includegraphics[width=0.5\textwidth]{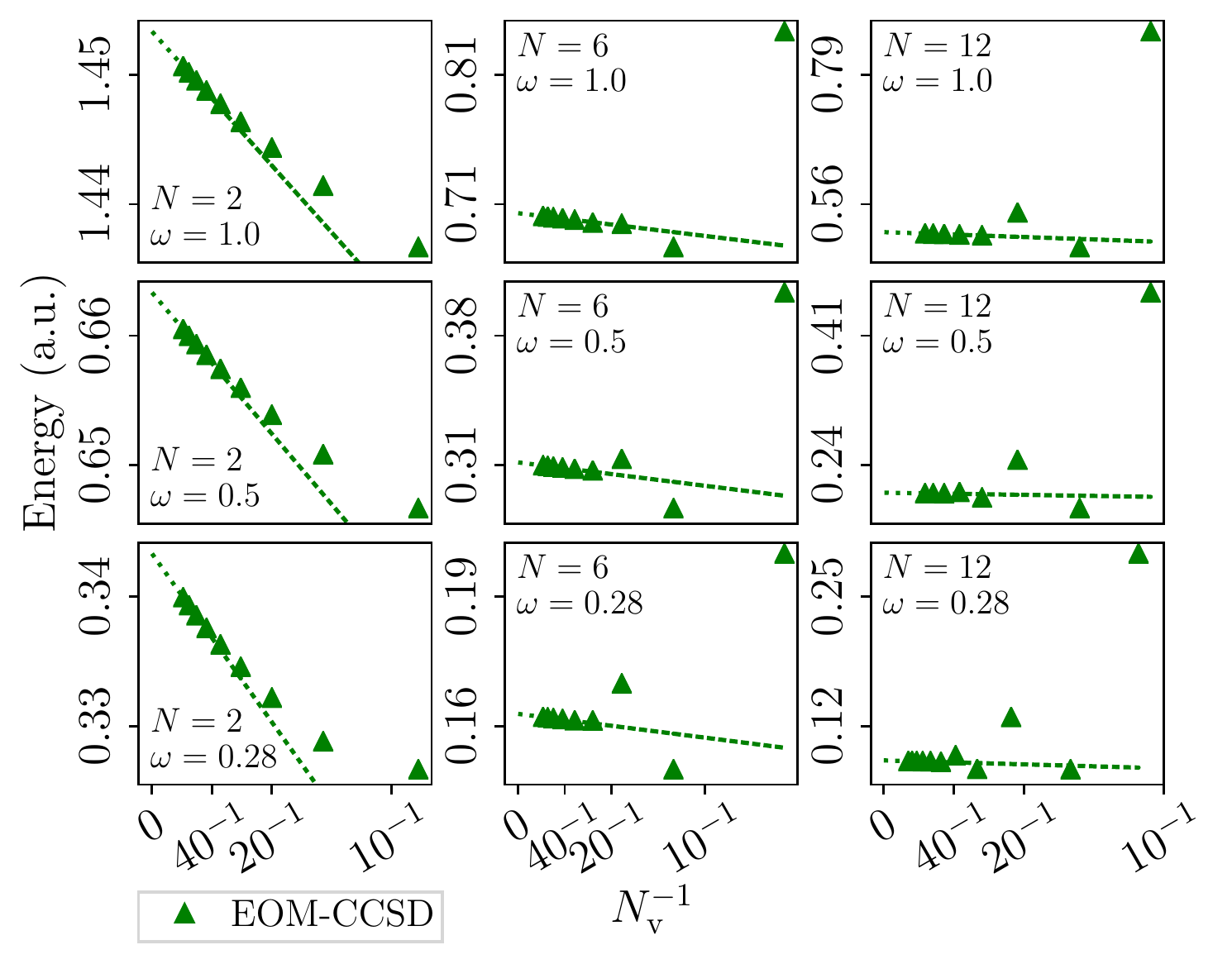}
\caption{%
  Third \gls{eeeomccsd} excitation energy for $N \in \{ 2, 6, 12 \}$
  electron systems with $\omega \in \{ 1.0, 0.5, 0.28\}$ retrieved as a
  function of the inverse number of virtual orbitals $N_{\mathrm{v}}^{-1}$.
  All energies are in Hartree.%
}\label{fig:eom_fig3}
\end{center}
\end{figure}

Analogously to the ground state, we need to converge the excitation energies
carefully with the basis set.  Figures \ref{fig:eom_fig}, \ref{fig:eom_fig2} 
and \ref{fig:eom_fig3} give evidence that the \gls{eeeomccsd} excitation 
energies converge in a similar manner to the complete basis set limit.
However, the slope is significantly less steep, resulting in excitation energies
with relatively small basis set incompleteness errors when employing $N_{\mathrm{v}}=77$.
Note that in the case of $N = 12$ and $\omega = 0.28$ we use $N_{\mathrm{v}}=114$ for the extrapolation.
We estimate the \gls{cbs} limit of the excitation energies using an identical
extrapolation procedure as outlined in the previous section.

\begin{figure}[htb]
\begin{center}
\includegraphics[width=0.5\textwidth]{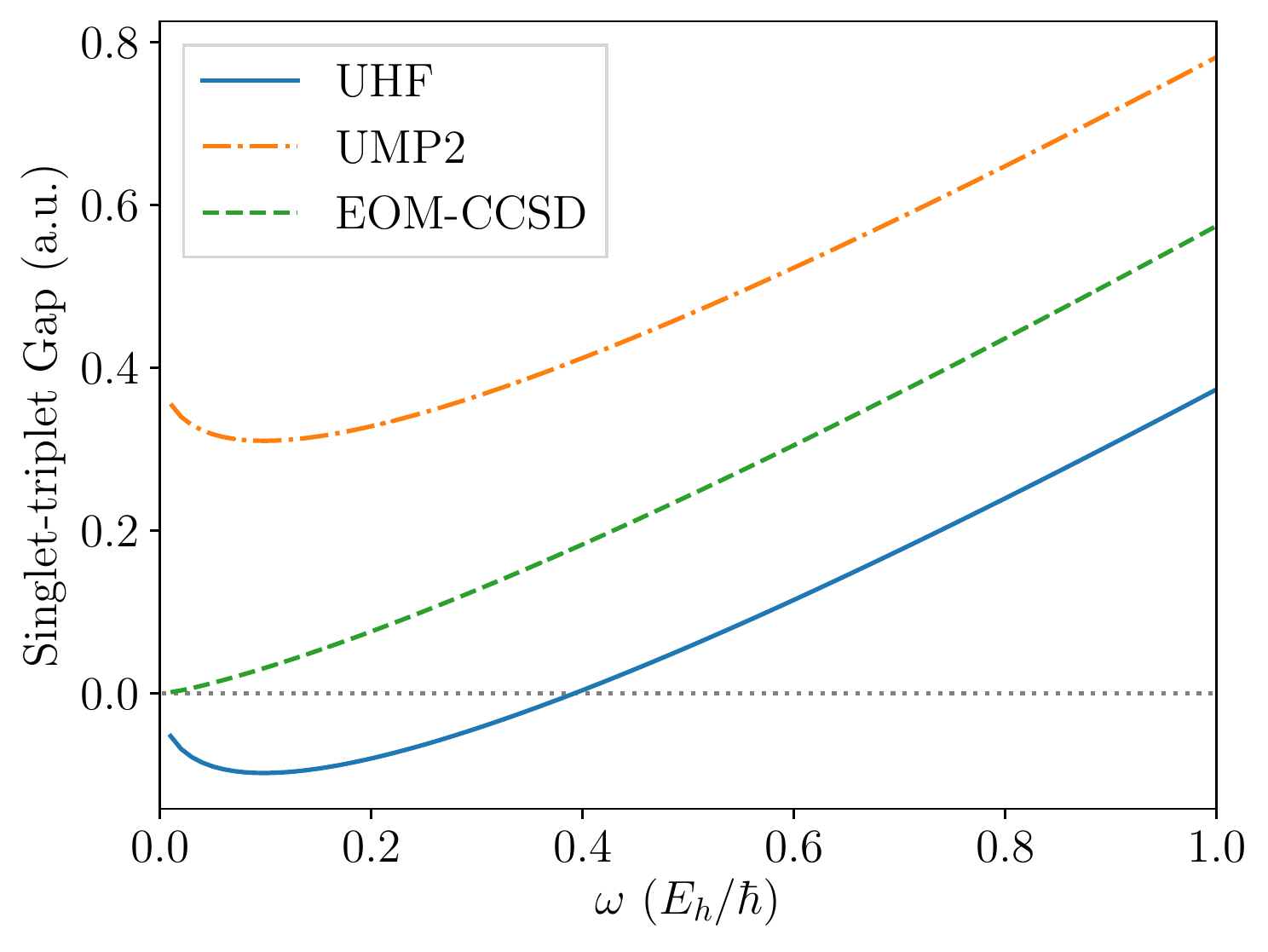}
\caption{%
  Singlet-triplet gap calculated with \gls{uhf}, \gls{ump2} and
  \gls{eeeomccsd} as a function of $\omega$ in Hartree. All calculations are done
  with $N_{\mathrm{v}}=10$.\label{fig:sing-trip}}
\end{center}
\end{figure}

Figure~\ref{fig:sing-trip} shows the first excitation energy (singlet-triplet gap) for $N=2$ as a function of $\omega$.
\gls{eeeomccsd} calculations predict an excitation energy that decreases with
decreasing $\omega$.
Approximating the singlet-triplet gap on the level of \gls{uhf} theory yields an inter system crossing at $\omega=0.3926$~a.u..
However, \gls{uhf} energies neglect correlation effects, which are expected to be
larger in magnitude for the singlet state than for the triplet state.
It has already been discussed that the singlet-triplet crossing predicted by \gls{uhf} results from the neglect of the electron-electron correlation~\cite{1.0}.
Indeed, we find that \gls{ump2} and \gls{eomcc} theory predict no singlet-triplet crossing.
Details on how the \gls{ump2} and \gls{uhf} singlet-triplet gap was calculated can be found in appendix~\ref{appendixc}.

Finally, Table~\ref{table2} summarizes the \gls{cbs} excitation energies from Fig. 
\ref{fig:eom_fig}, \ref{fig:eom_fig2} and \ref{fig:eom_fig3}. It shows that the 
linear scaling of the excitation energies with $\omega$ is qualitatively unchanged when comparing $N=2$, $N=6$ and $N=12$ electron systems.

\begin{table}
  \caption{%
    \gls{cbs} limit excitation energies for $N \in \{ 2, 6, 12 \}$ electron systems
    with $\omega \in \{1.0, 0.5, 0.28\}$.
    All quantities are expressed in atomic units.
    \label{table2}}
  \begin{tabularx}{0.5\textwidth}{XXXXX}
  \toprule
    $\omega$ (a.u.) & Electrons  &First excitation &Second excitation  &Third excitation  \\
  \midrule
    1.0 & 2  &0.5943  &0.9999  &1.4571\\
        & 6  &0.5218  &0.5532  &0.7028\\
        & 12 &0.4752  &0.4834  &0.5138\\
  \midrule
  0.5 & 2   &0.2530 &0.4999  &0.6609\\
      & 6   &0.2136 &0.2271  &0.3152\\
      & 12  &0.1913 &0.1951  &0.2063\\
  \midrule
  0.28 & 2   &0.1212 &0.2800  &0.3361\\
       & 6   &0.0991 &0.1018  &0.1589\\
       & 12  &0.0883 &0.0897  &0.0897\\
  \bottomrule
  \end{tabularx}
\end{table}

Our findings show that all excitation energies scale linearly with $\omega$.
Further we have compared our singlet excitation energies to values from Ref.~\cite{2.0} and they are
in excellent agreement as summarized in Table~\ref{tableEOMComp}.
The remaining differences of the excitation energies can be attributed to the \gls{cbs} extrapolation procedure and
also to the numerical procedures regarding the Coulomb integrals and the wave function, as described in the
previous section.

\begin{table}
  \caption{%
	  \gls{cbs} limit ground state and excitation energies for $N =  2$ electron systems with $\omega \in \{1.0, 0.5\}$ compared to coupled cluster energies from Ref.~\cite{2.0} on the right.
    All quantities are expressed in atomic units.
    \label{tableEOMComp}}
  \begin{tabularx}{0.5\textwidth}{XXXX}
  \toprule
    $\omega$ (a.u.) & Ground state  &Second excitation  &Third excitation  \\
  \midrule
    1.0 &3.002/3.003 &1.000/1.004 &1.457/1.456 \\
  \midrule
    0.5 &1.661/1.662 &0.500/0.502 &0.661/0.660\\
  \bottomrule
  \end{tabularx}
\end{table}

\section{Conclusion and Summary}

In this paper we have investigated a model Hamiltonian for two dimensional
\glspl{qd} using quantum chemical many-electron theories including HF, \gls{mp2}
and CCSD. For the study of excited states we have employed the
equation-of-motion formalism of CCSD theory (EOM-CCSD).  We have outlined a
numerical method to compute the Coulomb integrals for arbitrary orbitals
represented on a discrete numerical grid. Although this method is
computationally less efficient than recursive schemes for orbitals that
correspond to Gaussians or their derivatives, we note that it can become
potentially useful for different model Hamiltonians that include
a one-body part and corresponding eigenfunctions which are difficult to expand using
Gaussians or their derivatives.

We have investigated the convergence of the computed correlation energies
for ground and excited states with respect to the number of virtual orbitals
numerically, finding a convergence behavior for two dimensional \glspl{qd}
which is identical to the basis set
convergence of the second-order correlation energy in perturbation theory of the
three dimensional electron gas.  Furthermore, we have performed an analytic
derivation for the two dimensional \gls{qd} on the level of second-order
perturbation theory that supports this convergence behavior.  Based on this
analysis, we have extrapolated all computed correlation energies for ground and
excited states to the complete basis set limit assuming a $1/N_{\mathrm{v}}$
convergence of the remaining finite basis set errors.

The computed ground state energies in a range of $\omega =  0.28$~a.u., which
corresponds to a strongly correlated regime, to $\omega = 1.0$~a.u., has
revealed that the HF energy scales linearly with respect to $\omega$ and that the
relative contribution of the \gls{mp2} and CCSD correlation energies to the ground
state energy increases with decreasing $\omega$.  Furthermore, we have observed
that with decreasing $\omega$ the relative difference between the \gls{mp2} and
CCSD correlation energy is increasing, outlining that CCSD captures higher order
correlation effects than \gls{mp2}.

Using the \gls{eeeomccsd} formalism, we have calculated the first three excitation energies of the \gls{qd} and partly compared them to values from the literature.
Our findings show that the excitation energies scale linearly with $\omega$ and
for $N=12$ and $\omega=0.28$ the second and third excitation become numerically degenerate.

Finally, our work also demonstrates that two dimensional \gls{qd} model
Hamiltonians serves not only as a suitable tool for experimental \glspl{qd} but
can also be used as efficient and well-controlled testing ground of approximate
many-electron theories to study ground and excited state properties. Using a
single parameter to tune the confinement via the harmonic potential, the
Hamiltonian can be modified to switch between different regimes of correlation
strengths and investigate the accuracy of finite-order perturbation theories.
However, we find that EOM-CCSD performs qualitatively correctly for the
investigated parameter ranges and that the remaining errors are expected to be
only of quantitative interest. In future work we seek to investigate different
levels of EOM theories and compare to other widely-used electronic structure
theories that treat ground and excited state phenomena.

\section*{Acknowledgements}
The authors thankfully acknowledge support and funding from the European Research Council (ERC) under the European Unions Horizon 2020 research and innovation program (Grant Agreement No 715594).
The computational results presented have been achieved in part using
the Vienna Scientific Cluster (VSC).

\bibliography{Paper.bib}

\appendix

\section{Analytic Solution of the Integral over the 2D Coulomb Kernel}
\label{Analytic Solution}

To perform the integration in equation~\ref{Coulomb Integral} we discretize the
integration domain into hypercubes with an edge length of $\Delta x$ centered at
$x_i$, $x_j$, $y_k$, $y_l$.
\begin{widetext}
\begin{equation}\label{CI Discretized}
  \sum_{ijkl}
    \int%
      _{x_i - \frac{\Delta x}{2}}%
      ^{x_i + \frac{\Delta x}{2}}
    \int%
      _{x_j - \frac{\Delta x}{2}}%
      ^{x_j + \frac{\Delta x}{2}}
    \int%
      _{y_k - \frac{\Delta x}{2}}%
      ^{y_k + \frac{\Delta x}{2}}
    \int%
      _{y_l - \frac{\Delta x}{2}}%
      ^{y_l + \frac{\Delta x}{2}}
      \frac{%
        \psi_{mnopqrst}(x_1,y_1,x_2,y_2)
      }{%
        \sqrt{(x_1-x_2)^2 + (y_1-y_2)^2}
      }
      \mathrm{d}x_1
      \mathrm{d}x_2
      \mathrm{d}y_1
      \mathrm{d}y_2.
\end{equation}
\end{widetext}
With
  $ i,j,k,l \in \mathds{Z} $,
  $x_i = i\Delta x$,
  $x_j = j\Delta x$,
  $ y_k = k\Delta x$,
  $y_l = l\Delta x$
and
%\begin{widetext}
\begin{equation}
%\psi_{abcdefgh}(x_1,y_1,x_2,y_2) = \psi_{ab}^*(x_1,y_1) \psi_{cd}^*(x_2,y_2)
  %\psi_{ef}(x_1,y_1) \psi_{gh}(x_2,y_2).
\psi_{mnopqrst}
  = \psi_{mn}^*
    \psi_{op}^*
    \psi_{qr}
    \psi_{st}.
\end{equation}
%\end{widetext}
Employing simple quadrature, we approximate the wave function from~\ref{CI
Discretized} to be constant within each integration block.
\begin{widetext}
\begin{equation}\label{1. Approx}
  \sum_{ijkl}
    \psi_{mnopqrst}(x_i,y_j,x_k,y_l)
    \iiiint
    \frac{%
      \mathrm{d}x_1
      \mathrm{d}x_2
      \mathrm{d}y_1
      \mathrm{d}y_2
    }{%
      \sqrt{(x_1-x_2)^2 + (y_1-y_2)^2}
    }
\end{equation}
\end{widetext}
This leaves us with the integral over the Coulomb kernel, which cannot be
treated in the same manner due to points with $x_i=x_j$ and $y_k=y_l$, where the
Coulomb kernel becomes singular.
We solve this problem using the Laplace transformation of the Coulomb kernel,
leading to a simplified expression.
We start with
\begin{gather}\label{coulombbox}
  \int_{x_i - \frac{\Delta x}{2}}^{x_i + \frac{\Delta x}{2}}
  \int_{x_j - \frac{\Delta x}{2}}^{x_j + \frac{\Delta x}{2}}
  \int_{y_k - \frac{\Delta x}{2}}^{y_k + \frac{\Delta x}{2}}
  \int_{y_l - \frac{\Delta x}{2}}^{y_l + \frac{\Delta x}{2}}
  \frac{%
    \mathrm{d}x_1
    \mathrm{d}x_2
    \mathrm{d}y_1
    \mathrm{d}y_2
  }{%
    \sqrt{(x_1-x_2)^2 + (y_1-y_2)^2}
  }.
\end{gather}
Applying the Laplace transformation
\begin{gather}
  \frac{1}{|\vec{r_1}-\vec{r_2}|}
  =
  \frac{2}{\sqrt{\pi}}\int_0^{\infty}\mathrm{d}t e^{-t^2(|\vec{r_1}-\vec{r_2}|)^2}
\end{gather}
yields
\begin{widetext}
\begin{gather}
  \frac{2}{\sqrt{\pi}}
  \int_0^{\infty}\mathrm{d}t
  \int_{-\frac{\Delta x}{2}}^{\frac{\Delta x}{2}}
  \int%
      _{\Delta i \Delta x - \frac{\Delta x}{2}}%
      ^{\Delta i \Delta x + \frac{\Delta x}{2}}
    \mathrm{d}x_1
    \mathrm{d}x_2
    \ e^{-t^2(x_1-x_2)^2}
  \int_{-\frac{\Delta x}{2}}^{\frac{\Delta x}{2}}
  \int%
      _{\Delta j \Delta x - \frac{\Delta x}{2}}%
      ^{\Delta j \Delta x + \frac{\Delta x}{2}}
    \mathrm{d}y_1
    \mathrm{d}y_2
    \ e^{-t^2(y_1-y_2)^2}
\end{gather}
\end{widetext}
with $\Delta i = i - k$ and $\Delta j = j - l$.
The integration over $x_1$ and $x_2$
\begin{gather}
  \int_{-\frac{\Delta x}{2}}^{\frac{\Delta x}{2}}
  \int%
      _{\Delta i \Delta x - \frac{\Delta x}{2}}%
      ^{\Delta i \Delta x + \frac{\Delta x}{2}}%
  \mathrm{d}x_1
  \mathrm{d}x_2
  \ e^{-t^2(x_1-x_2)^2}
\end{gather}
can be done analytically using the error function, analogously for $y_1$ and
$y_2$.
The result is
\begin{widetext}
\begin{align*}
F(\Delta i, t)
  = \frac{1}{ 2 t^{2} }
  (   e^{-\Delta x^{2} (\Delta i - 1)^{2} t^{2}}
  - 2 e^{-\Delta x^{2}  \Delta i^{2} t^{2}}
  +   e^{-\Delta x^{2} (\Delta i + 1)^{2} t^{2}}
  + \Delta x \sqrt{\pi} t \\
    ( -2\Delta i     \mathrm{erf}(\Delta x \Delta i t)
    + (\Delta i + 1) \mathrm{erf}(\Delta x (\Delta i + 1) t)
    - (\Delta i - 1) \mathrm{erf}(\Delta x (-\Delta i + 1)t)
    )
  )
\end{align*}
\end{widetext}
And this leaves us with a 1D integral over the variable $t$ for every $\Delta i$
and $\Delta j$.
\begin{gather}
  a(\Delta i, \Delta j)
  = \frac{2}{\sqrt{\pi}}
    \int_0^{\infty}
    \mathrm{d}t F(\Delta i, t) F(\Delta j, t)
\end{gather}

\begin{figure}[htb]\label{fig6}
\begin{center}
\includegraphics[height=0.5\textwidth,width=0.5\textwidth,keepaspectratio]{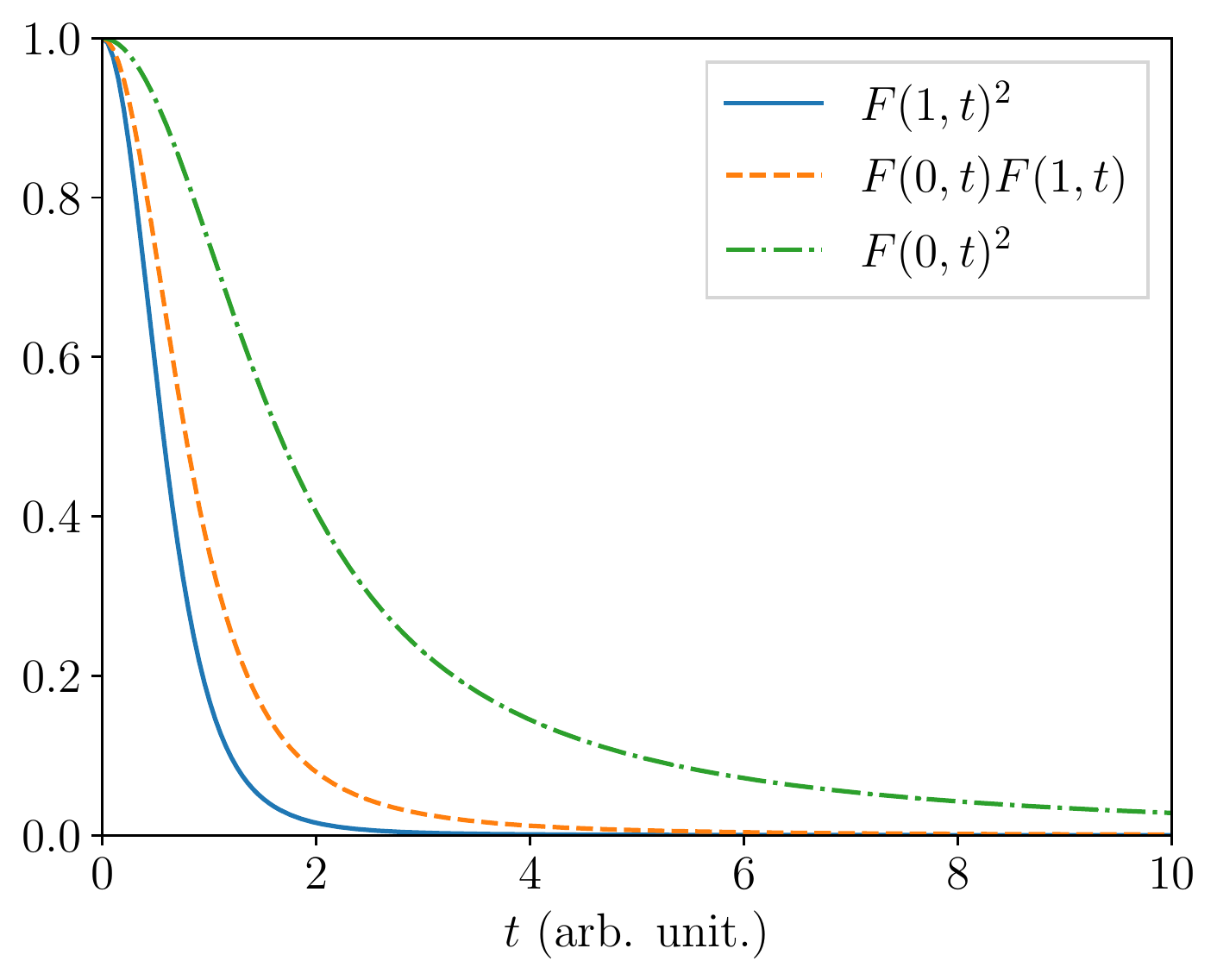}
  \caption{%
    $F(1, t)^2, F(0, t)F(1, t), F(0, t)^2$ from left to right. See main text for more details.
    \label{fig:plot_t}}
\end{center}
\end{figure}

As it can be seen in Figure~\ref{fig:plot_t} the integrand $F(\Delta i,
t)F(\Delta j, t)$ is well behaved and can be integrated numerically without much
computational cost.
In some special cases for example $\Delta i = \Delta j = 0$ the analytic
solution is available.
\begin{widetext}
\begin{gather}
  \int_{- \frac{\Delta x}{2}}^{\frac{\Delta x}{2}}
  \int_{- \frac{\Delta x}{2}}^{\frac{\Delta x}{2}}
  \int_{- \frac{\Delta x}{2}}^{\frac{\Delta x}{2}}
  \int_{- \frac{\Delta x}{2}}^{\frac{\Delta x}{2}}
    \frac{%
      \mathrm{d}x_1
      \mathrm{d}x_2
      \mathrm{d}y_1
      \mathrm{d}y_2
    }{%
      \sqrt{(x_1-x_2)^2 + (y_1-y_2)^2}
    }
  = -\frac{4}{3}
     \left(
       \sqrt{2}-1-3\mathrm{asinh}(1)
     \right)\Delta x^3
\end{gather}
\end{widetext}
But for the general case $\Delta i \neq \Delta j$ we have to solve the integral
numerically.

The functional form of the integral is not dependent on the domain of
integration.
Therefore the integral will always be proportional to $\Delta x^3$ times a
constant $a(\Delta i, \Delta j)$.
Note that the Constants $a(\Delta i, \Delta j)$ are not dependent on $\Delta x$.
Equation \ref{CI Discretized} can be rewritten as
\begin{gather}
  \sum_{ijkl}
    \psi_{mnopqrst}(x_i,y_j,x_k,y_l)
    a(\Delta i, \Delta j)
    \Delta x^3.
\end{gather}
We now evaluate the Coulomb integrals in real space numerically.
Note that the constants $a(\Delta i, \Delta j)$ only need to be computed once
and can be used for every $\Delta x$.
$\Delta i$ and $\Delta j$ define the distance of the integration region from the
singularity in steps of $\Delta x$. Approximating the Coulomb kernel by a
constant in the region
of integration becomes more accurate with increasing distance from the
singularity.
So a cutoff has to be chosen where the distance to the singularity is big enough
such that we can use the constant approximation.
With $\Delta i = 25$ and $\Delta j = 0$ equation \ref{coulombbox} with the
constant approximation gives $0.04 \Delta x^3$, while evaluated with our scheme
it gives $0.0400054 \Delta x^3$. Thus we have chosen $\Delta i = 25$ as cutoff.

Note that this evaluation scheme for the Coulomb integrals can be generalized to 
three dimensional systems straightforwardly.

\section{Asymptotic behavior of the Correlation Energy}\label{ABOTCE}

In order to extrapolate to the complete basis set limit of the correlation
energy, we need to derive an expression that yields that basis set truncation
error as a function of the number of virtual orbitals.
The correlation energy in second-order perturbation theory is given by
\begin{gather}
  E_{\mathrm{corr}}
    =
  \sum_{k}
    \frac{%
      |\bra{0}g_{ij}\ket{k}|^2
    }{%
      E_k-E_0
    }.
    \label{Correlation Energy}
\end{gather}
Where $\ket{0}$ denotes the ground state, $k$ is a excited state of the
unperturbed Hamiltonian and $E_0$ and $E_k$ are the corresponding energies.
In theory, the summation goes over all excited states but in practice we have to
truncate the summation at some cutoff $k_{\mathrm{cut}}$.
To replace the cutoff energy with the number of virtual orbitals in the above
equation, we have to employ equation~\ref{2D orbitals} and equation~\ref{2D
energy}.
We are only interested in the asymptotic behavior of the cutoff error, which is
defined by
\begin{equation}
  \label{ecorr asym}
  E_{\mathrm{err}}
  = \lim_{N_{\mathrm{v}} \to \infty}
      \sum_{k_{\mathrm{cut}}}^{\infty}
        \frac{|\bra{0}\frac{1}{r_{ij}}\ket{k}|^2}{E_k-E_0}.
\end{equation}
Furthermore, we can use the formula
\begin{gather}
  \lim_{n \to \infty} e^{-\frac{x^2}{2}}
  H_n(x) \sim \frac{2^n}{\sqrt{\pi}}
  \Gamma\left(
    \frac{n+1}{2}
  \right)
  \cos\left(
    x \sqrt{2n} - \frac{n \pi}{2}
  \right)
\end{gather}
to approximate our excited states with a simple $\cos$ function.
Now the Coulomb integral can be calculated analytically, which leaves us with
the result
\begin{gather}
  \label{coulomb asym}
  \lim_{R \to \infty}
  \bra{(00),(00)}
    \frac{1}{r_{ij}}
  \ket{(RR),(RR)}
  =
  \frac{%
    4^R \Gamma\left(\frac{1+R}{2}\right)^4
  }{%
    \pi^4 (R!)^2
  }
\end{gather}
where $R$ denotes the shell of the orbital.
Inserting this result into equation \ref{ecorr asym} and using the approximation
for the gamma function
\begin{gather}
  \lim_{x \to \infty} \Gamma(x+1)
  \sim
  \sqrt{2 \pi x}\left(\frac{x}{e}\right)^x
\end{gather}
gives us
\begin{gather}
  E_{\mathrm{err}}
  =
  \sum_{R}^{\infty}
    \frac{4(R-1)^2}{\pi^6 R^5}.
\end{gather}
In the above equation, the sum can be replaced by an integration, yielding
\begin{gather}
  E_{\mathrm{err}}
  =
  \frac{4}{\pi^6}
  \left(
    -\frac{1}{2R^2} + \frac{2}{3R^3} - \frac{1}{4R^4}
  \right).
\end{gather}
As the final step we have to convert the shell $R$ to the number of orbitals
$N_{\mathrm{v}}$.  By assuming filled shells we can write
\begin{gather}
  R
  =
  \frac{1+\sqrt{8N_{\mathrm{v}}+1}}{2}
\end{gather}
which gives us the final result for the basis set error of second-order
perturbation theory correlation energies computed using a truncated basis in the
limit of $N_{\mathrm{v}} \rightarrow \infty$:
\begin{gather}
  \lim_{N_{\mathrm{v}} \to \infty}
    E_{\mathrm{err}}
  \sim
  \frac{1}{N_{\mathrm{v}}}.
\end{gather}

\section{Singlet Triplet Gap calculation}\label{appendixc}
In order to calculate the singlet and triplet ground state energy of the 2
electron \gls{qd} with \gls{hf} and \gls{mp2} theory the following Slater determinants have been
used:
\begin{gather*}
	\ket{\Psi_{\mathrm{singlet}}} = \ket{(00,\uparrow)(00,\downarrow)} \\
	\ket{\Psi_{\mathrm{triplet}}} = \ket{(00,\uparrow)(01,\uparrow)}%
  .
\end{gather*}
The HF ground state energy for the 2 electron \gls{qd} is given by
\begin{gather*}
	E_{\mathrm{HF}} = \bra{\Psi_{\mathrm{gs}}}\hat{H}\ket{\Psi_{\mathrm{gs}}} + \bra{\Psi_{\mathrm{gs}}}\hat{V}\ket{\Psi_{\mathrm{gs}}}
\end{gather*}
where $\hat{H}$ is the single-body part of the Hamiltonian and $\hat{V}$ is the Coulomb repulsion between the electrons.
Inserting the ansatz for the wave functions of singlet and triplet states and applying the Slater-Condon
rules gives
\begin{gather*}
	E_{\mathrm{s}}= \omega + \braket{0000|0000} \\
	E_{\mathrm{t}}= 2\omega + \braket{0001|0001} - \braket{0000|0101}.
\end{gather*}
The \gls{mp2} ground state energy is
\begin{gather*}
	E_{\mathrm{MP2}} = E^{\mathrm{(0)}} + E^{\mathrm{(1)}} + E^{\mathrm{(2)}} \\
	E^{\mathrm{(0)}} = \bra{\Psi_{\mathrm{gs}}}\hat{H}\ket{\Psi_{\mathrm{gs}}} \\
	E^{\mathrm{(1)}} = \bra{\Psi_{\mathrm{gs}}}\hat{V}\ket{\Psi_{\mathrm{gs}}} \\
	E^{\mathrm{(2)}} = \sum_{k \neq \Psi_{\mathrm{gs}}} \frac{|\bra{k}\hat{V}\ket{\Psi_{\mathrm{gs}}}|^2}{E_{\mathrm{k}}-E_{\mathrm{gs}}} .
\end{gather*}
Applying the Slater-Condon rules and using the same singlet and triplet wave functions as for \gls{hf}
yields a additional contribution to the \gls{hf} energy
\begin{gather*}
	E_{\mathrm{MP2,s}}= E_{\mathrm{s}} + \sum_{\mathrm{abcd}} \frac{|\braket{0000|abcd}-\braket{00ab|00cd}|^2}{\omega(-a-b-c-d)} \\
	E_{\mathrm{MP2,t}}= E_{\mathrm{t}} + \sum_{\mathrm{abcd}} \frac{|\braket{0001|abcd}-\braket{00ab|01cd}|^2}{\omega(1-a-b-c-d)}
\end{gather*}

\end{document}